# Metalens-coupled terahertz NbN hot electron bolometer mixer

D. Ren[1], J. R. G. Silva[2], S. Cremasco[1], Z. Zhao[1], W. Ji[1], J. de Graaff[1], A. J. L. Adam[1], and J. R. Gao[1,2]*

[1]Department of Imaging Physics, Delft University of Technology, Lorentzweg 1, 2628 CJ, Delft, the Netherlands.
[2]Space Research Organization Netherlands (SRON), Niels Bohrweg 4, 2333 CA Leiden, and Landleven 12, 9747 AD Groningen, the Netherlands.
*Corresponding author (email: j.r.gao@sron.nl).

**Abstract:** Enabled by planarized phase engineering, metalenses based on metasurfaces offer compact and scalable solutions for applications such as sensing, imaging, and virtual reality. They are particularly attractive for multi-pixel, large-scale heterodyne focal plane arrays in space observatories, where a flat metalens array on a silicon wafer can replace individual lenses, greatly simplifying system integration and beam alignment. In this work, we demonstrate, for the first time, a superconducting niobium nitride (NbN) hot electron bolometer (HEB) mixer coupled with a silicon-based metalens operating at terahertz frequencies. The metalens phase profile was derived from a finite-size Gaussian beam source using the Rayleigh–Sommerfeld diffraction integral, and its focusing behavior was validated through 2D simulation. Experimentally, the metalens-coupled NbN HEB receiver exhibited a noise temperature of 1800 K at 1.63 THz. The power coupling efficiency from free space to the mixer via the metalens was measured to be 25%. Measured far-field beam profiles are Gaussian-like with sidelobes below −14 dB. These results demonstrate the feasibility of integrating metalenses with HEB mixers for THz detection, offering a scalable path for compact focal plane arrays in space-based THz instrumentation.

Superconducting hot electron bolometer (HEB) mixers are the most sensitive heterodyne detectors in the terahertz (THz) regime, typically covering frequencies from 1 to 6 THz.[1,2] HEB mixers based on niobium nitride (NbN) bridges have demonstrated near quantum-noise-limited performance,[3,4] enabling the detection of extremely weak signals and providing high spectral resolution with $f/\Delta f \geq 10^6$, where f is the frequency. Their ability to resolve fine-structure atomic and molecular lines, along with Doppler shifts in molecular clouds encoded in the frequency offset from the local oscillator (LO), makes them indispensable for high-resolution THz spectroscopy.

Heterodyne instruments based on HEB mixers are critical for space-based astronomy,[5] enabling the detection of ionized carbon ([CII] at 1.901 THz), atomic oxygen ([OI] at 4.745 THz), water, and other key tracers of cosmic chemistry. Single-pixel NbN HEBs or small arrays have been deployed in space on the ESA *Herschel* Space Observatory,[6] on an airborne platform of the *Stratospheric Observatory for Infrared Astronomy* (SOFIA),[7] and on NASA balloon-borne THz observatories, including the *Stratospheric Terahertz Observatory* (STO2)[8] and the *Galactic/Extragalactic ULDB Spectroscopic Terahertz Observatory* (GUSTO).[9]

A typical NbN bridge, with submicron dimensions (< 0.5 μm), requires efficient free-space coupling via either a lens-antenna[9] or feedhorn-waveguide[7] structure. In lens-antenna systems, an antireflection-coated elliptical silicon lens focuses a Gaussian beam into a diffraction-limited spot at the antenna, achieving a calculated coupling efficiency of 92% at 1.63 THz.[4]

The 8-pixel arrays used in *GUSTO*, assembled from individual elliptical lens–HEB mixer units into a common metal block, represent the current state-of-the-art.[9] Such arrays enable large-area sky surveys, as mapping speed scales inversely with pixel count (assuming constant per-pixel sensitivity). However, scaling beyond 8 pixels, for instance, toward 100, faces practical challenges due to the labor-intensive alignment and beam-pointing verification required for each chip at 4.2 K.[10]

We propose an alternative architecture for large HEB FPAs using flat silicon metalenses in place of elliptical lenses. In this approach, the FPA is formed by bonding two well-aligned silicon wafers: one with an array of metalenses, the other with antenna-connected HEBs. This requires only a single alignment step for the entire array. Unlike individually fabricated elliptical lenses, metalenses can be batch-fabricated using standard microfabrication processes, offering a compact, scalable, and cost-effective solution, similar to approaches pioneered for infrared FPAs.[11]

Since the introduction of metasurfaces for wavefront control,[12] metalenses based on multilayer plasmonic or all-dielectric metasurfaces have been demonstrated across a broad spectral range.[13–15] Numerous THz metalenses have been reported.[16–19] For example, an all-dielectric metalens composed of rectangular silicon pillars achieved a theoretical focusing efficiency of 61.3% and a measured efficiency of 45.8% at 1 THz.[19] Here, focusing efficiency refers to the fraction of incident power concentrated at the focal spot.

We adopted this all-silicon design for several reasons: (1) silicon exhibits low loss at THz frequencies, allowing efficient transmission and phase control; (2) it provides excellent thermal and mechanical compatibility with the HEB chip, minimizing interface losses, simplifying fabrication, and improving alignment precision.

To date, most THz metalens research has focused on free-space (air-to-air) light focusing. Although a silicon metalens integrated with an antenna-coupled transition-edge sensor bolometer has been proposed, it has only been explored through simulations.[20] To the best of our knowledge, no experimental demonstration of a metalens integrated with an antenna-coupled HEB has been reported.

In this work, we present the first experimental demonstration of a metalens coupled to a NbN HEB mixer operating at 1.63 THz. Using standard heterodyne techniques, we measured the receiver noise temperature and evaluated the optical coupling efficiency from the metalens to the mixer.

Figure 1a shows the schematic of the metalens-coupled HEB mixer. The structure consists of a Si micropillar-based metasurface on a thin Si substrate, a thick Si spacer, and a thin Si HEB chip, together forming a configuration resembling a solid-immersion lens. The spacer, in combination with the HEB chip substrate, extends the optical path length of the metasurface, allowing the beam to focus properly onto the spiral antenna integrated on the HEB chip. A superconducting NbN bolometer is positioned between the two antenna arms, as illustrated in Fig. 1b, with the focal spot designed to coincide with the antenna center.

For simplicity, we refer to the metasurface sample as the "metalens" throughout the paper. The metalens has a diameter of 7.5 mm and a designed focal length of 5115 µm in Si. These dimensions were chosen to fit an existing mixer block used for heterodyne measurements. The individual thicknesses are: 515 µm for the metalens, 4220 µm for the spacer, and 380 µm for the HEB chip.

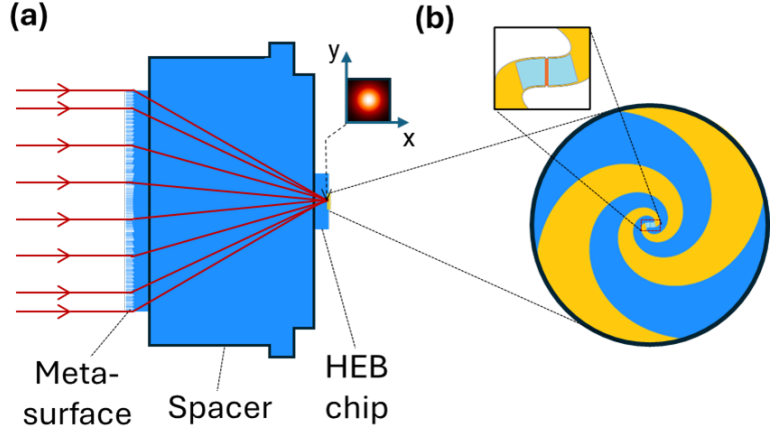

**Fig. 1.** (a) Schematic of a metalens-coupled HEB mixer, consisting of a metasurface, a disc-shaped spacer, and an HEB chip, all fabricated from silicon. The incident THz plane wave (red arrows) propagates toward the metasurface, which imparts spatially varying phase shifts across its aperture, converting the wave into a converging spherical wavefront focused onto the x–y plane where the HEB antenna is located. The focused beam is expected to be Gaussian with a beam waist of 100 µm, as illustrated. (b) Top view of the spiral antenna (yellow) on the HEB chip, with a zoomed-in view of the NbN microbridge at the center of the feed.

To derive the phase distribution of the electric field at the metalens plane, defined approximately at the interface between the metasurface and the Si substrate, where the incident beam undergoes phase modulation, we use the Rayleigh–Sommerfeld diffraction integral to back-propagate a narrow source from the designed focal plane (x–y plane), leveraging the reciprocity of a lens, as illustrated in Fig. 1a. The formula for the x-component (or y-component, depending on the spot polarization) of the electric field, u(x,y,z), at the metalens plane is given by[21]:

$$u(x,y,z) = -\frac{1}{2\pi} \iint_{-\infty}^{\infty} U(x_0, y_0) \cdot g(x - x_0, y - y_0, z, \lambda_0) dx_0 dy_0 \qquad (1)$$

where the propagation kernel $g(x,y,z,\lambda) = exp(iknR) \cdot (ikn - 1/R) \cdot z/R^2$, $R^2 = x^2 + y^2 + z^2$, $k = 2\pi/\lambda$, $n$ is the refraction index, $\lambda$ is the wavelength, and $U(x_0, y_0)$ is the electrical field at the focal plane.

For this proof-of-concept demonstration, we selected a Gaussian beam source with a 100 µm waist, rather than optimizing for maximum focusing efficiency or perfect beam-mode matching to the antenna. A narrower source was avoided because our metalens design has a high numerical aperture (NA) of 2. High-NA designs inherently lead to smaller focal spots but often suffer from lower focusing efficiency, producing broader, less efficient beams that deviate from the target beam. This is due to diffraction from the flat lens structure, which can redirect light into unwanted zero- and higher-order background modes.[22]

To compute the scalar electric field, we assume a refractive index n=3.384 for Si at 4 K,[23] since the lens operates at cryogenic temperatures, and a vacuum wavelength λ₀ to be 184.3 µm. Figure 2a shows the calculated phase distribution at the metalens plane, located 5020 µm from the HEB antenna. The phase profile exhibits a radially increasing gradient, indicating the need for finer spatial phase control near the lens edge to achieve accurate wavefront shaping.

To accommodate this high phase gradient, we adopt an all-Si metasurface approach following Ref. 19, using subwavelength meta-atoms with a 15 µm pitch—well below the wavelength in Si (~54.46 µm). The meta-atoms consist of square Si pillars[24] with varying widths, as illustrated schematically in Fig. 2b, to impart the required local phase shifts.

In principle, a constant pillar height could be assumed. However, in our metalens design, the pillar height varies with width due to fabrication constraints, as discussed in the fabrication section. Figure 2c shows this experimentally observed dependence, based on pre-fabrication tests of a similar metalens. The pillar height ranges from 100 to 150 µm as the width decreases from 13 to 5 µm.

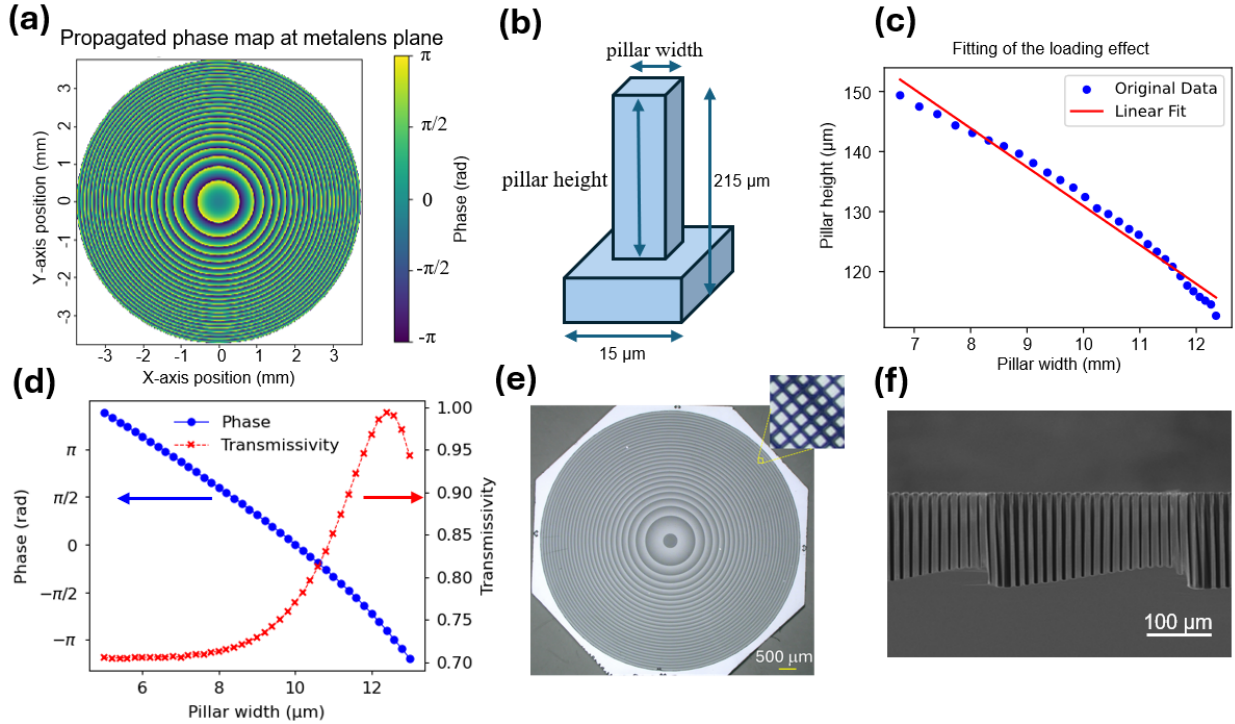

**Fig. 2.** (a) Simulated phase map at the metalens plane for focusing at 1.63 THz, showing concentric phase zones. (b) Schematic of a Si meta-atom unit with a lateral pitch of 15 µm and a total height of 215 µm, including both the pillar and part of the substrate. The pillar width is varied, which also affects the achievable height, as shown in (c). (c) Measured pillar height as a function of width. (d) Simulated phase shift (blue circles) and transmitted intensity versus square pillar width. (e) Optical micrograph of a fabricated Si metalens (with a slightly different phase map than in (a)); inset: zoomed-in view. (f) SEM cross-sectional image showing high-aspect-ratio Si pillars.

We employed the eigenmode solver in COMSOL Multiphysics to calculate the phase shift and transmission of meta-atoms for incident light at 1.63 THz, by varying the pillar width from 5 to 13 µm. The corresponding pillar height was also varied, following the fitted linear relationship between width and height shown in Fig. 2c. Figure 2d presents the calculated phase shift as a function of width, ranging from 1.4π to -1.2π, thereby achieving full 0 to 2π phase coverage. The simulated transmission varies between 70% and 100%. The COMSOL simulations were performed using periodic boundary conditions with high-order diffraction modes enabled, allowing us to capture the grating-like behavior inherent to the periodic metalens structure.[17] Based on these simulations, a meta-atom library was established to fulfill the required phase distribution of the metalens.

The design of the metalens then becomes straightforward. Specifically, the detailed layout of the metasurface is realized by assigning, at each location in the phase map (Fig. 2a), a pillar from the library that provides the required phase shift. This one-to-one correspondence enables a direct translation of the desired phase distribution into a physical layout of square pillars with varying widths.

The focusing behavior of the designed metalens was simulated in two dimensions and validated using COMSOL Multiphysics. The result is shown in Fig. 3, with additional simulation details provided in the Supplementary Material.

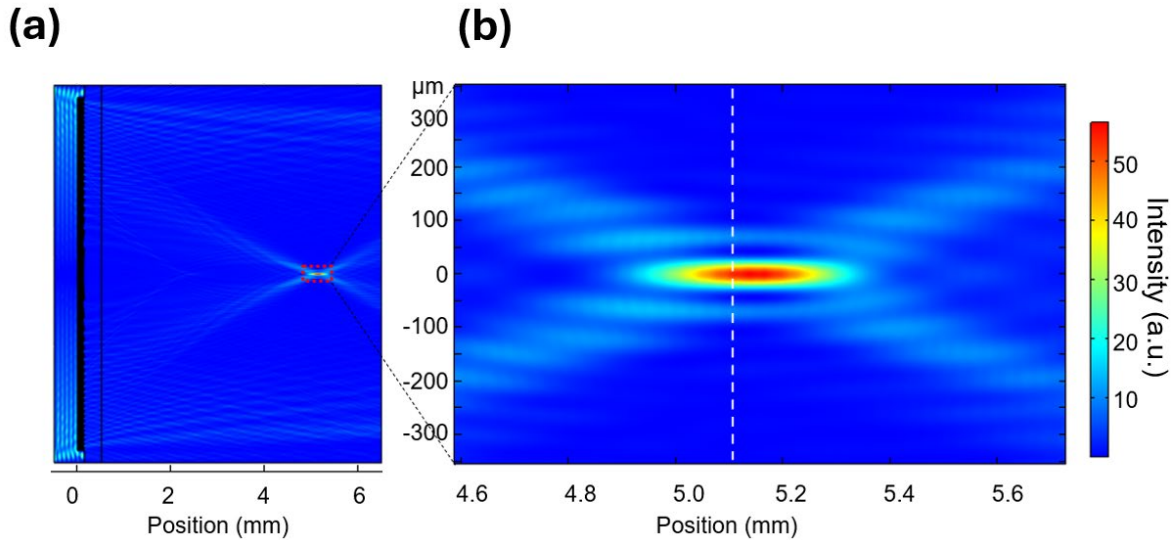

**Fig. 3.** Electric field intensity distribution of the metalens at 1.63 THz, obtained using a modified 2D simulation approach. (a) Cross-sectional view along the optical axis, showing clear focusing behavior with a well-defined focal spot. (b) Zoomed-in view of the focal region from (a). The white dashed line indicates the position of the antenna (~5.115 mm).

The metalens was fabricated on a double-side polished silicon wafer (515 µm thick). All silicon components, including the spacer and HEB chip, were high-resistivity (≥ 5 kΩ·cm). To fabricate the metalens, a laser writer was used for lithography to define the etching mask pattern of square pillars. Deep Reactive Ion Etching (DRIE) using a Bosch process in an ICP-RIE system was then applied to etch the silicon and form pillars with heights ranging from 110 to 141 µm. The variation in pillar height arises from the loading effect during the etching process,[25] as documented in Fig. 2c. A uniform pillar height could be achieved by introducing an etch-stop layer.

Figure 2e shows an optical micrograph of the fabricated Si metalens, with the inset providing a zoomed-in view. The pillar sizes and their periodic arrangement match the design, demonstrating successful pattern transfer with high spatial fidelity. Figure 2f presents a cross-sectional SEM image, revealing well-defined, high-aspect-ratio pillars and clearly illustrating the loading effect.

We used the same NbN HEB chip as in Ref. 4. To realize the metalens-coupled HEB mixer (in Fig. 1), a disc-shaped Si spacer, comparable in surface quality and dimensional accuracy to a conventional Si lens,[10] was employed. The spacer is mounted in the same mixer block used for elliptical Si lenses[4] and is designed for cryogenic operation. The spacer is necessary because the focal length exceeds the thicknesses of both the metalens wafer and the HEB chip.

The metalens sample (DMLv9_03) was aligned and glued to the front surface of the spacer (Φ = 10 mm), while the HEB chip was attached to the back surface (Φ = 8.9 mm). The alignment accuracy of both components relative to the spacer's optical axis was within 3 μm.

To evaluate the metalens performance, we conducted heterodyne measurements to determine the double-sideband (DSB) receiver noise temperature ($T_{rec}^{DSB}$) of the mixer, which is a figure of merit for a heterodyne receiver. The measurements were performed at 4.4 K using a 1.63 THz line from a far-infrared gas laser as the local oscillator (LO). A hot (295 K) and cold (77 K) blackbody load was alternately placed in front of the receiver. The experimental setup was identical to that used in our previous lens-HEB mixer study.[4]

The THz signal from the blackbody passed through air, a 3 μm Mylar beam splitter, an ultra-high molecular weight polyethylene window at 300 K, a 4 K heat filter (QMC), and finally the metalens, which focused the beam onto the spiral antenna. The resulting intermediate frequency (IF) signal, produced by mixing the blackbody radiation and LO, was routed through a circulator, amplified by a low-noise SiGe amplifier (LNA) at 4.2 K, filtered at 1.7 GHz with an 80 MHz bandwidth, and further amplified by a room-temperature LNA to produce the receiver output power.

The receiver output power, $P_{hot}$ and $P_{cold}$, measured when the mixer is exposed to hot and cold blackbody loads, respectively, is recorded as a function of HEB current (by varying the LO power) at an optimal bias voltage of 0.6 mV, as shown in Fig. 4a. Using the standard Y-factor method and the Callen–Welton formula for effective blackbody temperatures,[1,2] $T_{rec}^{DSB}$ is derived from the measured $P_{hot}$ and $P_{cold}$, and also plotted in the same figure.

The minimum $T_{rec}^{DSB}$ is 1794 K, recorded at a mixer current of 41.5 μA. Notably, this value already surpasses the performance reported in Ref. 26, which marked one of the earliest demonstrations of conventional lens-antenna NbN HEB mixers. The measured $T_{rec}^{DSB}$ includes contributions from the mixer itself, as well as optical losses and noise introduced by components in front of the mixer. These contributions are described by Eq. (1), derived from Ref. 2,5:

$$T_{rec}^{DSB} = T_{opt} + \frac{T_{mixer}^{DSB}}{G_{opt}G_{lens}} + \frac{T_{IF}}{G_{opt}G_{lens}G_{mixer}^{DSB}}, \qquad (1)$$

where $T_{opt}$ is the noise temperature contribution from the optical components in front of the metalens ($= 101\ K$), $T_{mixer}^{DSB}$ the mixer noise temperature, $G_{mixer}^{DSB}$ the mixer gain, $G_{opt}$ the total optical gain of the components before the lens, $G_{lens}$ the optical gain of the metalens or the power coupling efficiency, and $T_{IF}$ the noise temperature of the IF LNA chain ($= 6.5\ K$). The measured $T_{rec}^{DSB}$ can thus be used to extract the optical gain of the metalens.

To estimate $G_{lens}$, we use Eq.1 along with the following parameters from Ref. 4: $T_{mixer}^{DSB}$=221 K and $G_{mixer}^{DSB}$=-4.05 dB. These values were obtained using the same HEB chip mounted behind an antireflection-coated Si lens at 1.63 THz, with the lens contribution removed. We also adopt $G_{opt}$=-2.25 dB from Ref. 4, as the measurement setup was identical. From this, we determine $G_{lens}$=−6.3 dB, corresponding to a power coupling efficiency of approximately 23%.

A more direct and reliable estimation of $G_{lens}$ can be obtained by measuring the total receiver gain $G_{total}$ using the U-factor method.[28] The U-factor is defined as the ratio of the receiver output power at optimal operating conditions (where $T_{rec}^{DSB}$ = 1794 K) to the output power when the HEB is in the superconducting state and illuminated by the hot load. $G_{total}$ is then derived from the U-factor using the following expression (Eq. 2), as given in Ref. 27:

$$G_{total} = \frac{U(T_{IF}+T_{ref})}{T_{rec}^{DSB}+T_{293K}}. \qquad (2)$$

where $T_{ref}$ equals 4.2 K, which is the physical temperature of the 50 Ω resistor terminated to the circulator.

Using the measured U-factor of 10.48 dB and Eq. (2), $G_{total}$ is calculated to be –12.35 dB. Since $G_{total} = G_{opt} G_{lens} G_{mixer}^{DSB}$, we derive $G_{lens}$= -6.05 dB (or 25%) using the same $G_{mixer}^{DSB}$ and $G_{opt}$ values as before (from Eq. 1). Both methods yield consistent values for $G_{lens}$, confirming the reliability of our measurements.

Using the heterodyne beam measurement setup described in Ref. 28, we measured the far-field beam pattern of the mixer in the x–y plane at a distance of 450 mm from the metasurface. As shown in Fig. 4b, the beam exhibits a Gaussian-like profile. No sidelobes were observed above −14 dB relative to the main peak, limited by the signal-to-noise ratio of the measurement system, which is approximately −14 dB. The absence of sidelobes confirms the good Gaussian character of the metalens–antenna HEB mixer. The 2D inset in Fig. 4b shows an elliptical beam profile, attributed to the asymmetric geometry of the spiral antenna.

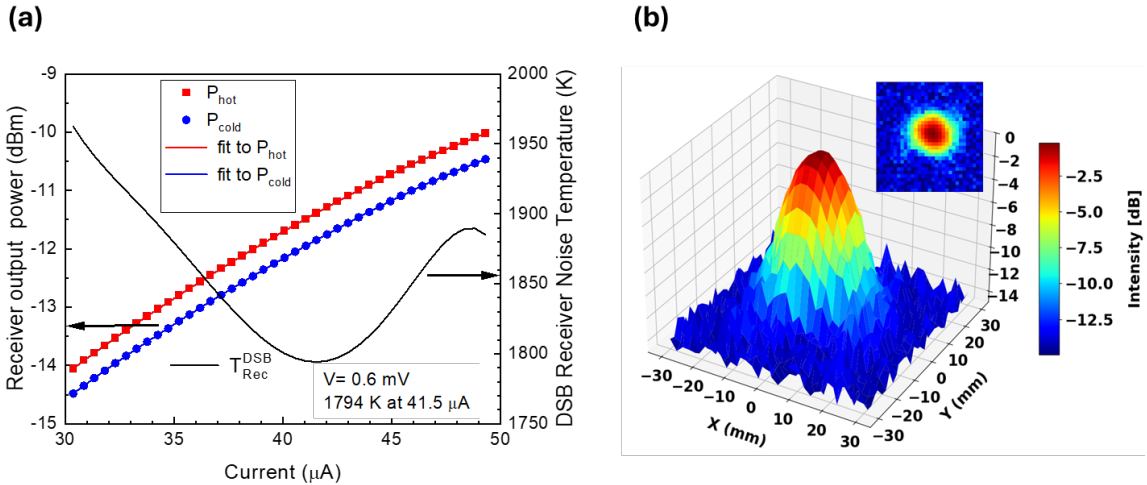

**Fig.4.** (a) DSB Receiver noise temperature $T_{rec}^{DSB}$ of the metalens-coupled HEB mixer measured at 4.4 K and 1.63 THz. The values are extracted from measured receiver output power versus HEB bias current in response to a hot load ($P_{hot}$) or a cold load ($P_{cold}$). Both measured data and fitted curves are shown. (b) Far-field beam profile of the metalens-coupled HEB mixer, measured at a distance of 450 mm from the metasurface. The 3D profile is shown on a dB scale; the corresponding 2D beam pattern is shown in the inset.

We now discuss the implications of the measured optical gain of the metalens. $G_{lens}$ can be decomposed into two components: the focusing efficiency of the metalens and the mode-matching efficiency between the focused beam and the antenna. Prior studies report focusing efficiencies of ~46%[19] or 60%[17] for all-Si metalenses, assuming no reflection losses at the air–Si interface. Given our measured $G_{lens}$ of 25%, the beam-to-antenna mode-matching efficiency is estimated to be no higher than ~54%.

To improve this, a new metalens design could adopt a phase profile derived from a Gaussian source with a beam waist of ~39 µm, consistent with simulated focusing from an elliptical Si lens.[29] Ultimately, an optimal solution would involve generating the phase profile directly from the beam of the spiral antenna used in the mixer.

To realize the desired phase profiles, smaller meta-atoms could be used to reproduce rapidly varying phase gradients, especially near the lens edges. However, losses from unwanted diffraction orders fundamentally limit beam narrowing and focusing efficiency.[22] Recent work has shown that inverse-designed pillar shapes can overcome this, achieving focusing efficiencies up to ~90% at infrared wavelengths.[30] This method allows direct control over structural complexity via Fourier decomposition of the surface gradient. Additionally, an antireflection structure could be introduced to reduce the ~15% reflection loss at the air–Si metasurface interface.[31] The ultimate goal is to raise $G_{lens}$ above 70%, which would substantially enhance the observing efficiency of an array relative to a single-pixel receiver.[5]

In summary, we demonstrate an all-Si metalens that couples free-space THz radiation to a spiral-antenna-coupled NbN HEB mixer at 1.63 THz. The receiver achieves a competitive DSB receiver noise temperature of 1800 K and exhibits a Gaussian-like far-field beam with sidelobes below −14 dB. Using standard heterodyne techniques, we accurately determine the metalens power coupling efficiency to be 25%. Pathways for further improvement are identified. This metalens–antenna mixer serves as a promising precursor for scalable HEB focal plane arrays in future space observatories.

Acknowledgment: This work was supported by the Horizon Europe Radioblocks project (Grant No. 101093934) and the TU Delft Space Institute. The authors thank Behnam Mirzaei for fabricating the HEB, Qing Yu for sharing expertise on metalens design, Shahab Dabironezare for valuable discussions on the antenna, Vishal Anvekar for the optical micrograph of the metalens, and Wouter Laauwen and Willem Jan Vreeling for their support with the measurements.

The data that supports the findings of this study are available from the corresponding author upon reasonable request.

# Supplementary Material

To validate the focusing behavior of the metalens, a full 3D COMSOL Multiphysics simulation could, in principle, be used to visualize the electric field distribution along the propagation path of the THz beam. However, due to the large computational domain required, such a simulation is not feasible. Instead, a 2D simulation was performed using a cross-sectional profile of the structure. In this model, the metasurface pillars are represented as square elements, with their widths adjusted to reproduce the phase shifts of the corresponding 3D pillars. The effective pillar width used in the 2D simulation, denoted as $P_{eff\_2D}$, was computed using the following relation:

$$P_{eff\_2D} = \left(\frac{P_{3D}}{Pitch}\right)^2 \cdot Pitch \tag{S1}$$

where $P_{3D}$ is the width of a pillar in the 3D metalens, and *Pitch* is the unit cell size. This adjustment ensures that the accumulated phase shift in the 2D model replicates the same phase modulation profile as in the full 3D metalens design.

To properly simulate the focusing behavior of the 3D metalens using a 2D representation, it is essential to account for the spatial dimensional mismatch. The 2D cross-sectional model inherently loses information in the direction perpendicular to the plane. In the full 3D system, the lens is radially symmetric, and the incident optical power is distributed over a circular area, where the differential area element scales as $dA \propto r dr$. In contrast, the 2D simulation represents a cylindrical slice and does not naturally incorporate this radial dependence.

To correct for this, the amplitude of the incident field was modulated as a function of the radial coordinate *r*, ensuring proper energy distribution across the lens radius. Specifically, the incident field amplitude was scaled as:

$$E_{in}(r) \propto \sqrt{r} \tag{S2}$$

To compensate for the absence of radial area scaling in the 2D simulation, an intensity weighting was applied to better approximate the energy convergence and focusing characteristics expected in a full 3D metalens. Figure 3 (in the manuscript) shows the simulated 2D intensity profile of the focused beam in silicon, obtained by illuminating the metalens from the left with a 1.63 THz plane wave incident on the pillar array. The metalens surface, defined as the interface between air and the tops of the pillars, is set as the zero point in the coordinate system.

A distinct focal spot is observed in Fig. 3a (in the manuscript) and becomes more evident in the zoomed-in view of Fig. 3b, where the beam focuses between 5100 µm and 5200 µm from the metalens surface. The beam waist at focus is approximately 25 µm, significantly smaller than the 100 µm input beam waist. This discrepancy likely reflects an inherent limitation of 2D simulations compared to full 3D modeling.[1]

---

[1] Y. Yang, J. Cheng, X. Dong, F. Fan, X. Wang, and S. Chang, "3D high-NA metalenses enabled by efficient 2D optimization," Optics Communications **520**, 128448 (2022).